\documentclass[10pt,letterpaper]{article}
\usepackage{opex3}
\usepackage{color}
\usepackage{cite}
\usepackage{braket}
\begin{document}

\title{A compact iodine-stabilized laser operating at 531 nm with stability at the $10^{-12}$ level and using a coin-sized laser module}

\author{Takumi Kobayashi$^{1,2,3*}$, Daisuke Akamatsu$^{1}$, Kazumoto Hosaka$^{1,3}$, Hajime Inaba$^{1,3}$, Sho Okubo$^{1,3}$, Takehiko Tanabe$^{1}$, Masami Yasuda$^{1}$, Atsushi Onae$^{1,3}$, and Feng-Lei Hong$^{1,2,3}$}

\address{$^1$National Metrology Institute of Japan (NMIJ), National Institute of Advanced Industrial Science and Technology (AIST), 1-1-1 Umezono, Tsukuba, Ibaraki 305-8563, Japan\\
$^2$Department of Physics, Graduate School of Engineering, Yokohama National University, 79-5 Tokiwadai, Hodogaya-ku, Yokohama 240-8501, Japan\\
$^3$JST, ERATO, MINOSHIMA Intelligent Optical Synthesizer Project, ERATO, 1-1-1 Umezono, Tsukuba, Ibaraki 305-8563, Japan}

\email{$^*$takumi-kobayashi@aist.go.jp} 



\begin{abstract}
We demonstrate a compact iodine-stabilized laser operating at 531 nm using a coin-sized light source consisting of a 1062-nm distributed-feedback diode laser and a frequency-doubling element. A hyperfine transition of molecular iodine is observed using the light source with saturated absorption spectroscopy. The light source is frequency stabilized to the observed iodine transition and achieves frequency stability at the $10^{-12}$ level. The absolute frequency of the compact laser stabilized to the $a_{1}$ hyperfine component of the $R(36)32-0$ transition is determined as $564\,074\,632\,419(8)$ kHz with a relative uncertainty of $1.4\times10^{-11}$. The iodine-stabilized laser can be used for various applications including interferometric measurements.
\end{abstract}

\ocis{(120.4800) Optical standards and testing; (120.3940) Metrology; (300.6320) Spectroscopy, high-resolution.} 


\section{Introduction}
Developments in optical frequency standards have been of great interest in a wide range of fields over many years and recently led to accuracies at the $10^{-18}$ level \cite{Bloom2014,Ushijima,Nicholson2015}. 
Iodine-stabilized lasers were one of the earliest optical frequency standards and are still important for practical applications due to their simplicity and robustness. Of these lasers, a frequency-doubled Nd:YAG laser, which is stabilized to an iodine absorption line at 532 nm, has attracted considerable attention owing to its very high frequency stability \cite{Eickhoff1995,Hall1999,Hong2004,Zang2007}. This is mostly due to the fact that molecular iodine has strong and narrow absorption lines near 532 nm. 
Applications using the Nd:YAG laser include a flywheel oscillator or an absolute frequency marker for an optical frequency comb \cite{Ye2001,Takamoto2005,Hong2005,Cingoz2012}, interferometric measurements of gauge blocks \cite{Bitou2003}, gravitational wave detection \cite{Musha2000,Kokuyama2010,Argence2010}, and a laser strainmeter for observing earth tides and earthquakes \cite{Takemoto2004}.

A compact iodine-stabilized laser emitting in the 532-nm region is attractive for some of the above applications. Some compact Nd:YAG systems have already been developed in previous experiments \cite{Bitou2003,Kokuyama2010,Argence2010,Hong2001}, one of which achieved a size of 30 × 45 cm$^{2}$ \cite{Hong2001}. A more compact configuration could be constructed by using a small iodine cell and light source. Recently, an iodine-loaded hollow-core photonic crystal fiber has been demonstrated as a potential alternative to an iodine cell \cite{Lurie2012}. Regarding the light source, a diode laser operating near 1064 nm can be a compact and low-cost alternative to the Nd:YAG laser. The linewidth of a diode laser, e.g., a distributed-feedback (DFB) diode laser, usually exceeds several hundred kilohertz, whereas that of a Nd:YAG laser is several kilohertz. Nevertheless, a diode-laser-based scheme may be suitable for applications that require extremely small systems, e.g., space applications \cite{Kokuyama2010,Argence2010}, and applications that do not demand very high frequency stabilities. 

This paper presents a compact iodine-stabilized laser emitting at 531 nm using second harmonic generation (SHG) of a 1062-nm DFB diode laser. This wavelength is determined by that of a commercially available diode laser. To our knowledge, this is the first report of an iodine-stabilized diode laser in the wavelength region near 532 nm. To construct a very simple laser system, we employ the following methods: (i) we adopt a recently-developed coin-sized module containing both the diode laser and an SHG element; (ii) we use a short iodine cell operated at room temperature; and (iii) we stabilize the laser frequency to a sub-Doppler resolution iodine line by the third-harmonic technique with the frequency modulation of the diode laser, which enables us to eliminate the need for the electro-optic modulator used in the frequency modulation sideband or modulation transfer methods \cite{Eickhoff1995,Hall1999,Hong2004,Zang2007}. All the optical parts of the iodine-stabilized laser are arranged on a $20\times30$ cm$^{2}$ breadboard, but can potentially fit into an area of considerably less than $10\times10$ cm$^{2}$. The above simplifications could degrade the frequency stability compared with that of previous Nd:YAG systems \cite{Eickhoff1995,Hall1999,Hong2004,Zang2007}. Here we demonstrate that our compact laser can achieve a frequency stability at the 10$^{-12}$ level and a frequency uncertainty at the $10^{-11}$ level, which are sufficient for various applications including gauge block measurements \cite{Bitou2003}.

\section{Experimental setup}
\label{experimentalmethodsection}
\begin{figure}[h]
\begin{center}
\includegraphics[width=11cm,bb=0 50 910 716]{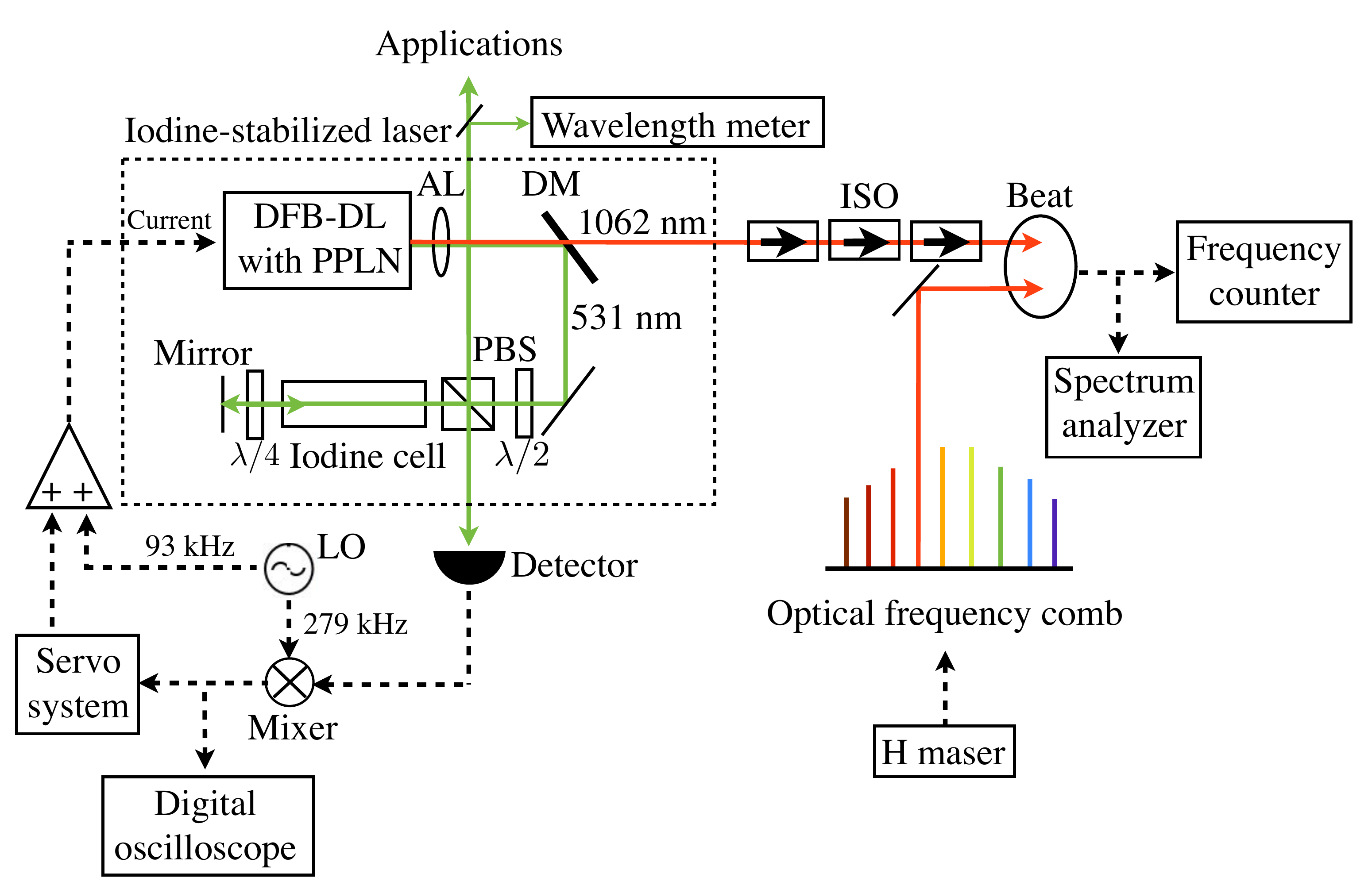}
\end{center}
\caption{Schematic diagram of the experimental setup (not to scale). DFB-DL: Distributed-feedback diode laser, PPLN: Periodically poled lithium niobate, AL: Aspheric lens, DM: Dichroic mirror, ISO: Isolator, PBS: Polarization beam splitter, LO: Local oscillator.}
\label{experimentalsetup}
\end{figure}
Figure \ref{experimentalsetup} is a schematic diagram of the experimental setup. The light source at 531 nm is a compact laser module (QDLASER, QLD0593) consisting of a DFB diode laser at 1062 nm, a semiconductor optical amplifier, and a periodically poled lithium niobate (PPLN) for SHG. The dimensions of the compact module are 21.9 mm $\times$ 5.6 mm $\times$ 3.8 mm (length $\times$ width $\times$ thickness). We mounted the module on a Peltier device, which we used to stabilize and change the temperature of the module. 

To perform Doppler-free spectroscopy of molecular iodine using the 531-nm laser module, we developed a simple and compact iodine spectrometer using saturated absorption. The 531-nm laser beam with a power of about 20 mW was collimated by an aspheric lens and separated from the fundamental light emitting at 1062 nm using a dichroic mirror. A half-wave plate $(\lambda/2)$ and a polarization beam splitter (PBS) were used to divide the 531-nm beam into two parts. One portion of this beam with a maximum power of $5$ mW was used as the output beam for possible applications and laser wavelength measurements. The other portion was sent to the saturated absorption spectrometer using a 6-cm-long iodine cell. The beam transmitted through the iodine cell was reflected back to the cell by a mirror. The original and back-reflected beams served as pump and probe beams, respectively, in the saturation spectroscopy. Since the beam passed through a quarter wave plate $(\lambda/4)$ twice, the polarization of the beam was rotated by 90$^{\circ}$. The resulting s-polarized probe beam was steered onto a photo detector by the PBS. To detect the Doppler-free signals, the frequency of the light source was modulated by adding a sinusoidal signal at 93 kHz to the injection current of the DFB diode laser. This modulation was detected from the probe beam by using the photo detector and demodulated by mixing the detected signal with the signal from a local oscillator (LO). To eliminate the Doppler background of the iodine absorption line, a third derivative signal was obtained by demodulating the detected signal at 279 kHz. This signal was fed back to the injection current of the DFB diode laser through a servo system for frequency stabilization.

The fundamental 1062-nm beam was sent to an optical frequency comb via three isolators with an isolation of 100 dB. The frequency comb was based on a mode-locked erbium-doped fiber laser operated at a repetition rate of $f_{\mathrm{rep}}=107$ MHz \cite{Inaba2006,Nakajima2010,Iwakuni2012}, which was self-referenced and phase locked to a hydrogen maser (H maser). The frequency of the H maser was calibrated against the Coordinated Universal Time (UTC) of the National Metrology Institute of Japan (NMIJ). A heterodyne beat note was detected between the 1062-nm light and a comb component, and measured by using a spectrum analyzer and a frequency counter. 

\section{Experimental results}
\label{experimentalresultsection}
\begin{figure}[h]
\begin{center}
\includegraphics[width=11cm,bb=0 80 1010 616]{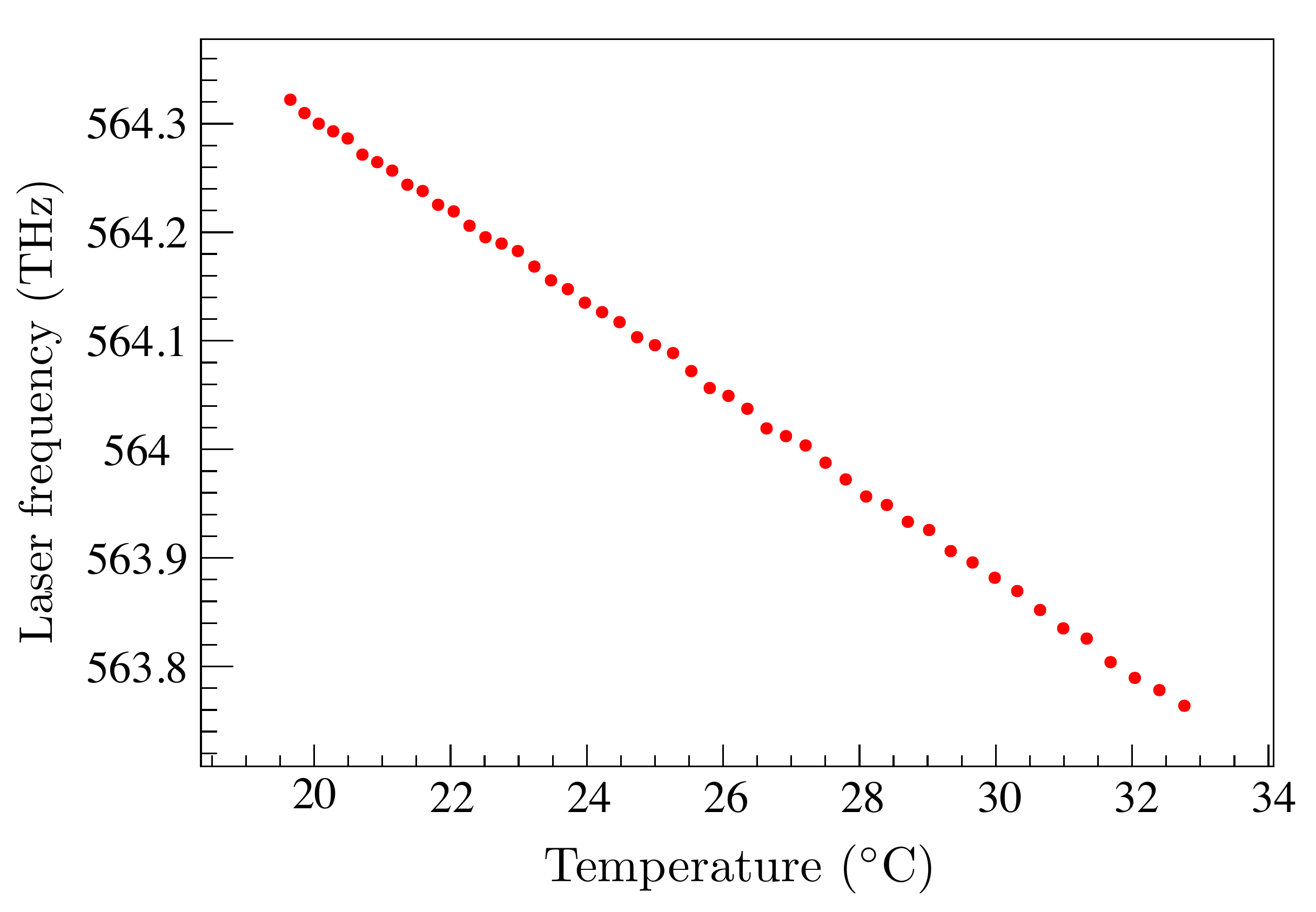}
\end{center}
\caption{Measured frequency tunability against the temperature of the compact laser module.}
\label{tunability}
\end{figure}
\begin{figure}[h]
\begin{center}
\includegraphics[width=11cm,bb=0 70 1010 716]{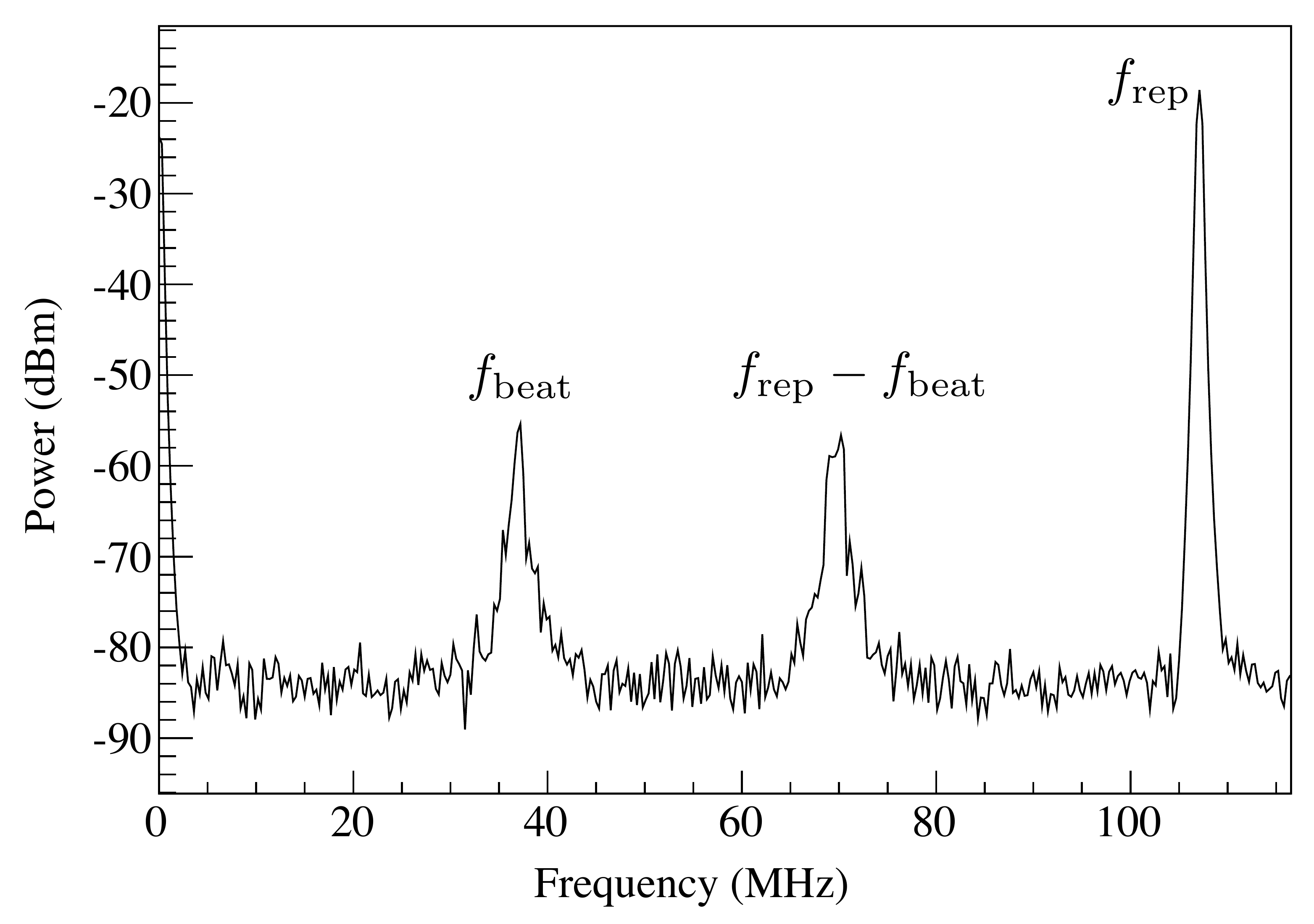}
\end{center}
\caption{Spectrum of the beat signal between the free-running compact laser and the optical frequency comb stabilized to the H maser. The resolution bandwidth was 300 kHz. $f_{\mathrm{beat}}$ denotes the beat frequency between the laser and the comb, and $f_{\mathrm{rep}}$ the repetition rate of the comb.} 
\label{beatspectrum}
\end{figure}
\begin{figure}[h]
\begin{center}
\includegraphics[width=11cm,bb=0 60 1010 716]{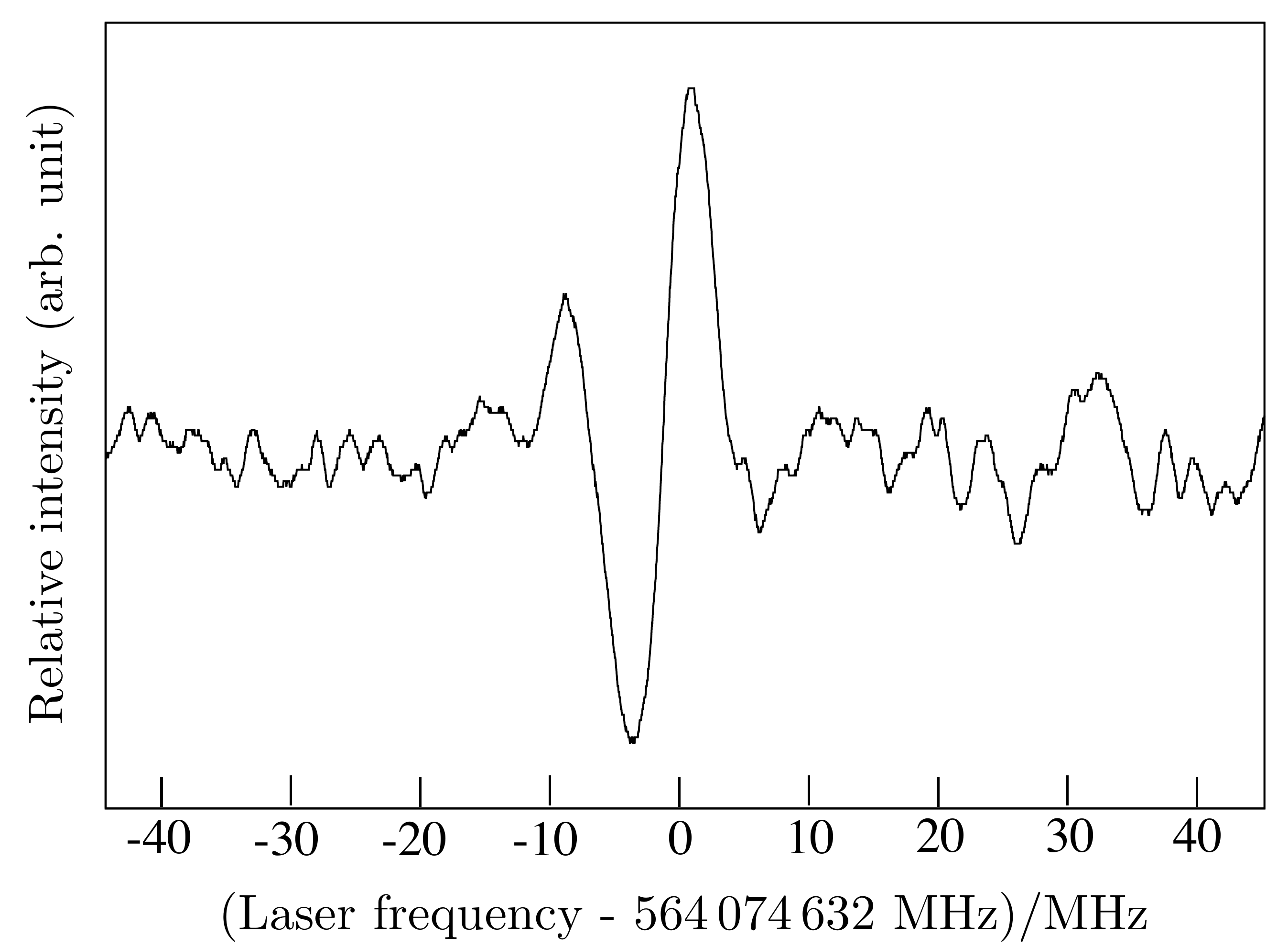}
\end{center}
\caption{Third derivative signal of the saturated absorption spectrum for the $a_{1}$ hyperfine component of the $R(36)32-0$ transition.}
\label{thirdderivativesignal}
\end{figure}

We studied the frequency characteristics of the compact laser module. Figure \ref{tunability} shows the laser frequency (converted from the wavelength) measured by a wavelength meter as a function of the module temperature. In these measurements, the injection current of the DFB diode laser was fixed at a constant value. The laser frequency was tunable over 500 GHz without mode hopping simply by changing the module temperature. We observed frequency instabilities at some temperatures during the measurements. However, single frequency laser oscillation was obtained at most temperatures. The laser power variation was less than $10\%$ across the 500-GHz tuning range.

Figure \ref{beatspectrum} shows the beat signal measured by the spectrum analyzer between the free-running compact laser and the comb stabilized to the H maser. The full width at half maximum (FWHM) of the beat note was $1-2$ MHz. This was mainly contributed from the fundamental 1062-nm laser linewidth, since the linewidth of the comb is expected to be $<10$ kHz\cite{Inaba2013}. Thus, the linewidth of the SHG light at 531 nm is estimated to be $2-4$ MHz. The beat signal was also used to measure the width of the frequency modulation of the laser light, when a modulation signal was added to the laser injection current. We observed a significant reduction of the modulation width at modulation frequencies above 300 kHz. This indicates that the maximum response rate of the laser frequency to the change in the injection current is about 300 kHz.

Figure \ref{thirdderivativesignal} shows the third derivative signal of the saturated absorption for the $a_{1}$ hyperfine component of the $R(36)32-0$ transition that we observed using our iodine spectrometer. This signal was obtained by tuning the injection current of the laser and recording the waveform of the signal with a digital oscilloscope (see Fig. \ref{experimentalsetup}). The temperature of the laser module was stabilized at about 25 $^{\circ}$C. The power of the pump beam (i.e., the incident laser power just before the iodine cell) was $P_{\mathrm{pump}}=12.7$ mW. The diameter $d$ of the beam inside the cell was $\sim1$ mm. The cold finger temperature was held at $T=25$ $^{\circ}$C, corresponding to an iodine pressure of $p=41$ Pa. Due to the absorption of the laser beam inside the cell, the laser power was reduced to $<1$ mW before the photo detector. The modulation width was set at $\Omega=12$ MHz (peak-to-peak). The FWHM of the observed spectrum in Fig. \ref{thirdderivativesignal} was about 8 MHz. The signal-to-noise (SN) ratio of the spectrum was approximately 10 in a bandwidth of 1 kHz. The short-term frequency stability of the iodine-stabilized light source is basically determined by the spectral linewidth and the SN ratio. From the observed spectrum, the frequency stability was estimated to be $1\times10^{-9}$ at an averaging time of $\tau=1$ ms.

\begin{figure}[h]
\begin{center}
\includegraphics[width=10cm,bb=0 50 1010 716]{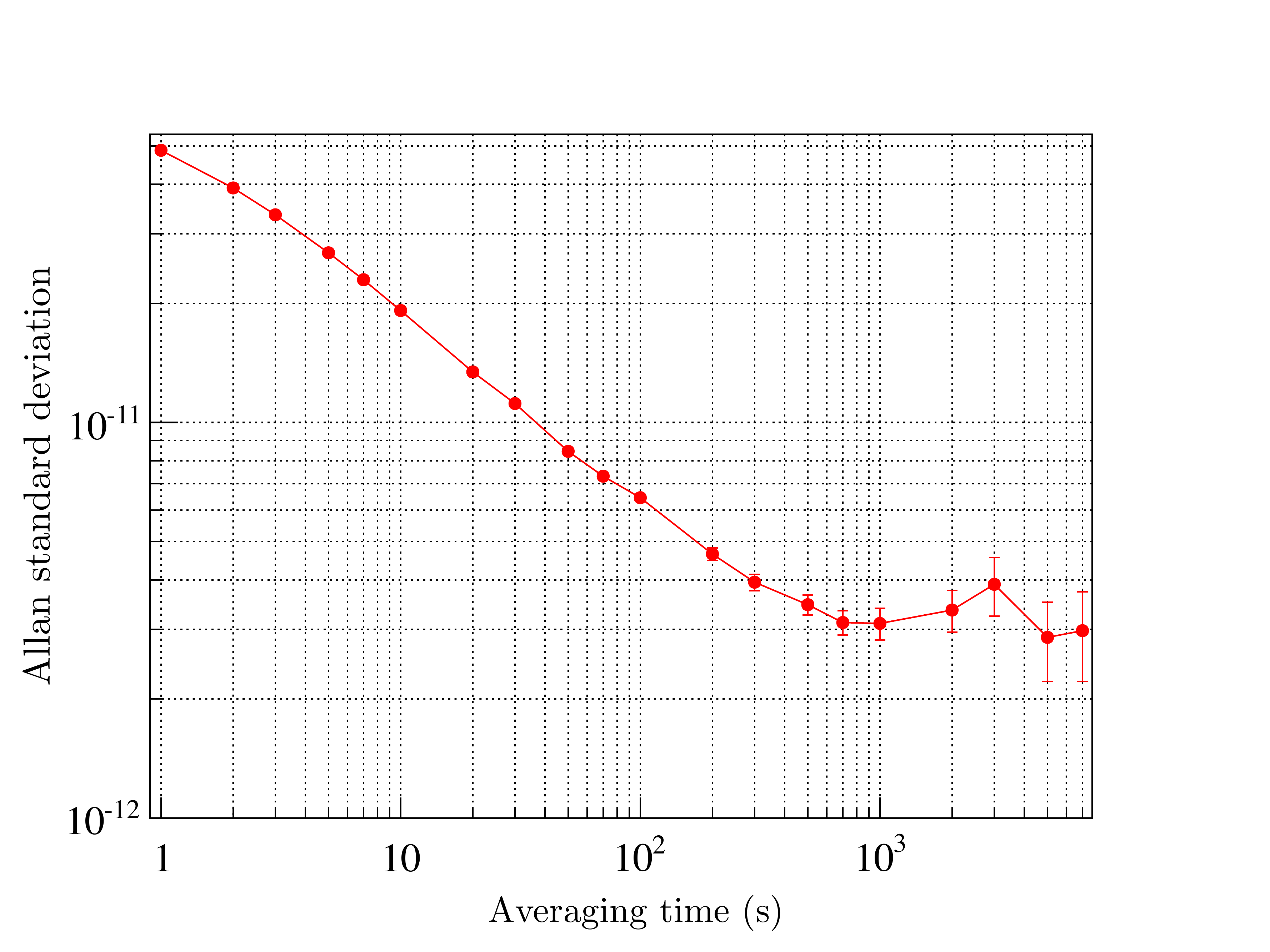}
\end{center}
\caption{Allan standard deviation calculated from the measured beat frequency between the compact laser locked on the $a_{1}$ component of the $R(36)32-0$ transition and the optical frequency comb stabilized to the H maser.}
\label{allandeviation}
\end{figure}
\begin{figure}[h]
\begin{center}
\includegraphics[width=10cm,bb=0 90 1010 716]{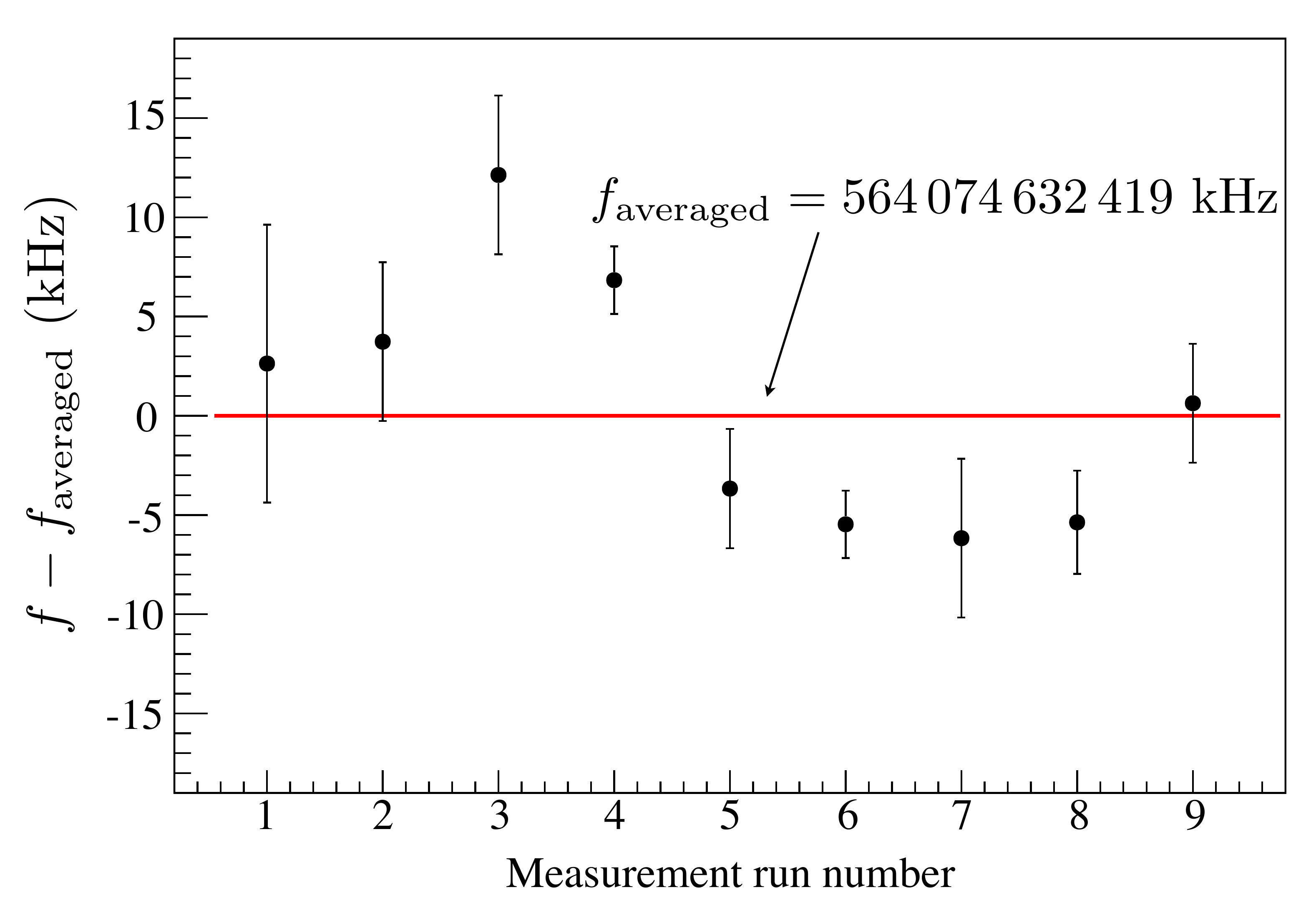}
\end{center}
\caption{Measured absolute frequency of the compact laser locked on the $a_{1}$ component of the $R(36)32-0$ transition. The solid red line indicates the weighted mean of the nine measured frequencies.}
\label{repeatability}
\end{figure}

Frequency stabilization was realized by the feedback control of the diode current using the observed third derivative signal. Figure \ref{allandeviation} shows the Allan standard deviation calculated from the measured beat frequency between the compact laser locked on the $a_{1}$ component of the $R(36)32-0$ transition and the comb stabilized to the H maser. The frequency counter used in the measurement was a $\Pi$-type counter. The Allan deviation of the H maser starts from the $10^{-13}$ level at 1 s, which is much smaller than that of the compact laser. Therefore, the observed stability of the beat frequency was mainly limited by the frequency stability of the compact laser. The frequency stability of the laser was found to be $5\times10^{-11}$ at $\tau=1$ s and reached $3\times10^{-12}$ at $\tau=700$ s. For $\tau>700$ s, the long-term stability was limited by a flicker floor at $3\times10^{-12}$. When we assume a $1/\tau^{1/2}$ slope, the observed stability at $\tau=1$ s is consistent with the short-term stability estimated from the spectral linewidth and SN ratio.

Since the frequency comb is referenced to UTC(NMIJ), i.e., the national frequency standard, the absolute frequency of the iodine-stabilized laser is obtained from the beat measurement. Figure \ref{repeatability} shows the results of nine measurements for the $a_{1}$ component of the $R(36)32-0$ transition obtained over several days. In these measurements, the experimental parameters were fixed such that $P_{\mathrm{pump}}=12.7$ mW, $d\sim1$ mm, $T=25$ $^{\circ}$C ($p=41$ Pa), and $\Omega=12$ MHz. Each measurement in Fig. \ref{repeatability} was calculated from more than 1000 beat frequency data, where each frequency datum was measured by a frequency counter with a gate time of 1 s. The uncertainty bar was given by the Allan standard deviation at the longest averaging time. For example, the uncertainty bar of 1.7 kHz for the measurement run number 6 in Fig. \ref{repeatability} was calculated using the Allan standard deviation of $3\times10^{-12}$ at $\tau=7000$ s in Fig. \ref{allandeviation}. The weighted mean of the nine measured frequencies in Fig. \ref{repeatability} was $564\,074\,632\,419$ kHz. 
\begin{figure}[h]
\begin{center}
\includegraphics[width=13cm,bb=0 20 1010 816]{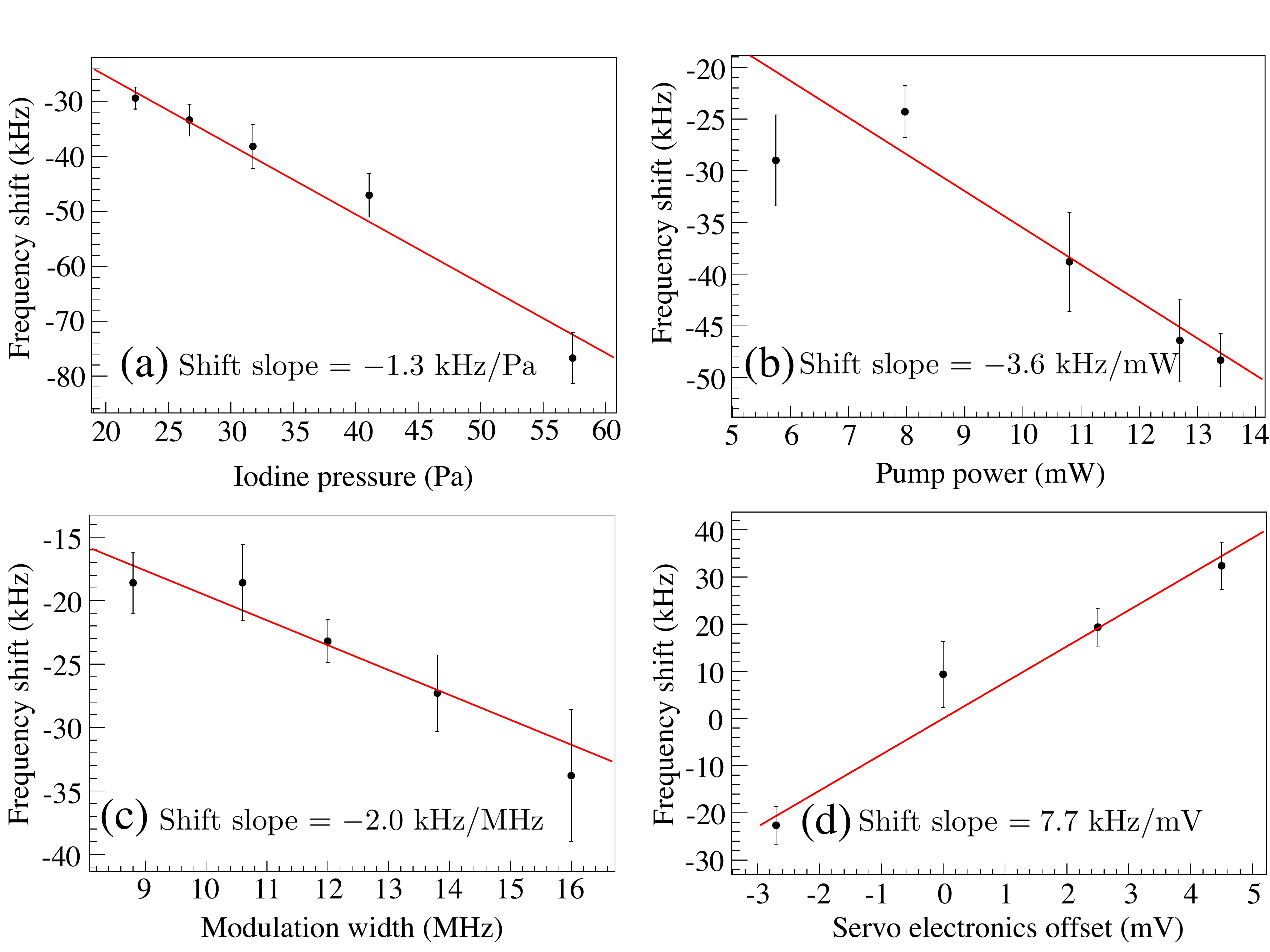}
\end{center}
\caption{Measured frequency shifts of the compact laser locked on the $a_{1}$ component of the $R(36)32-0$ transition: (a) pressure shift, (b) power shift, (c) modulation-induced shift, and (d) shift due to the servo electronics offset. The solid red line in each figure indicates the best fit of a linear function using the weighted least square method.}
\label{pressureshift}
\end{figure}

\begin{table}[h]
\caption{Most significant contributions to the estimated frequency uncertainty of the compact laser locked on the $a_{1}$ component of the $R(36)32-0$ transition.}  
	\label{uncertaintyesti}
	\begin{center} 
\begin{tabular}{lll}
\hline
\hline
Effect & Sensitivity & Contribution\\
\hline
Pressure shift & $-1.3$ kHz/Pa & $<0.3$ kHz \\
Power shift & $-3.6$ kHz/mW & $<4$ kHz \\
Modulation-induced frequency shift & $-2.0$ kHz/MHz & $<2$ kHz \\
Servo electronics offset & $7.7$ kHz/mV & $<4$ kHz\\
Cell impurity & & $5$ kHz\\
\hline
Statistics & & $2.1$ kHz \\
\hline
Total uncertainty & & $<8$ kHz\\
Relative uncertainty & & $1.4\times10^{-11}$\\
\hline
\hline
\end{tabular}
\end{center}
\end{table}

Several systematic frequency shifts of the compact iodine-stabilized laser were investigated. Figure \ref{pressureshift} (a) shows the measured pressure shift of the laser locked on the $a_{1}$ component of the $R(36)32-0$ transition. The measured slope of the pressure shift was $-1.3$ kHz/Pa. Although the cold finger temperature was stabilized within 10 mK, it is difficult to measure the temperature of the solid-state iodine crystal in the cold finger, which determines the iodine pressure \cite{Hong2004}. The uncertainty in the determination of the solid-state iodine was estimated to be $<0.5$ K. This corresponds to a pressure uncertainty of $<0.2$ Pa, resulting in a frequency uncertainty of $<0.3$ kHz. Figure \ref{pressureshift} (b) shows the measured frequency shift of the compact laser as a function of the pump power. The measured slope of the power shift was $-3.6$ kHz/mW. On the assumption that the uncertainty in the determination of the laser power is $<10\%$, the power shift effect gives rise to a frequency uncertainty of $<4$ kHz. It is known that the frequency modulation of the light source causes a significant frequency shift \cite{Robertsson2001}. Figure \ref{pressureshift} (c) shows the measured frequency shift induced by the modulation. The measured slope of the modulation-induced frequency shift was $-2.0$ kHz/MHz. The uncertainty in the determination of the modulation width was estimated to be $<1$ MHz, which results in a frequency uncertainty of $<2$ kHz. The servo electronics offset, i.e., a dc voltage offset between the baseline of the spectrum and the lock point, can shift the laser frequency. Figure \ref{pressureshift} (d) shows the measured frequency shift caused by the servo electronic offset. In the present experiment, the measured slope of this effect was $7.7$ kHz/mV. We adjusted the servo electronics offset to less than $0.5$ mV, which results in a frequency uncertainty of $<4$ kHz. Contamination in the iodine cell is known to cause a frequency shift. Previous studies \cite{Eickhoff1995} have shown that this effect can add an uncertainty of 5 kHz to measurement results.   

The most significant contributions to the estimated frequency uncertainty of the compact iodine-stabilized laser are summarized in Table \ref{uncertaintyesti}. During the experiment, the uncertainty of the H maser was confirmed to be less than $1\times10^{-14}$, corresponding to an absolute uncertainty of 0.006 kHz in the iodine transition frequency. The total uncertainty was estimated to be 8 kHz (relatively $1.4\times10^{-11}$), including the statistical uncertainty of 2.1 kHz obtained from Fig. \ref{repeatability}. The statistical uncertainty was inflated by the square root of the reduced chi-squared $\sqrt{\chi^{2}_{\mathrm{reduced}}}=2.3$. The absolute frequency of the compact laser stabilized to the $a_{1}$ component of the $R(36)32-0$ transition was determined as $564\,074\,632\,419(8)$ kHz with the experimental parameters $P_{\mathrm{pump}}=12.7$ mW, $d\sim1$ mm, $T=25$ $^{\circ}$C ($p=41$ Pa), and $\Omega=12$ MHz.

\section{Discussion}
The observed spectral linewidth of 8 MHz in Fig. \ref{thirdderivativesignal} is comparable to the expected value of $\Delta\nu_{\mathrm{th}}=10-12$ MHz, which was estimated from the pressure width, the power broadening, and the laser linewidth ($\Delta\nu_{\mathrm{laser}}=2-4$ MHz). The pressure width at $p=41$ Pa was tentatively estimated to be $\gamma_{\mathrm{p}}=6$ MHz using a pressure broadening coefficient of $\alpha=148$ kHz/Pa for the $R(36)32-0$ transition at 532 nm \cite{Eickhoff1995}. Since the spatial distribution of the laser power inside the iodine cell was not uniform due to the strong absorption, we used an average laser power of $P_{\mathrm{ave}}\sim6$ mW to calculate the Rabi frequency $x_{\mathrm{R}}$. On the assumption that the electronic transition dipole moment of the observed 531-nm transition is $\mu_{\mathrm{e}}=1$ D \cite{Lamrini1994} and the Franck-Condon factor $\left|\Braket{v^{'}=32,J^{'}=37|v^{''}=0,J^{''}=36}\right|^{2}$ (i.e., the square of the overlap integral between the ground and excited rovibrational wave functions) is 0.03 \cite{Klug2000}, $x_{\mathrm{R}}$ was estimated to be 3 MHz. The spectral linewidth $\Delta \nu_{\mathrm{th}}$ was then calculated from the relationship $\Delta \nu_{\mathrm{th}}=\sqrt{\gamma_{\mathrm{p}}^2+(2x_{\mathrm{R}})^2}+\Delta\nu_{\mathrm{laser}}$. We note that only order estimation was performed here, since our calculation was based on our limited knowledge of the parameters such as $\alpha$, $P_{\mathrm{ave}}$, and $\mu_{\mathrm{e}}$.

The observed frequency stability of our compact iodine-stabilized laser (shown in Fig. \ref{allandeviation}) was worse than that of a previously reported Nd:YAG laser locked on the $R(56)32-0$ transition (better than the $10^{-13}$ level) \cite{Eickhoff1995,Hall1999,Hong2004,Zang2007}. This is due to the relatively large spectral linewidth and low SN ratio observed in the present experiment. Since we used a 6-cm-long iodine cell, compared with iodine cells of 30 to 200 cm used in the previous experiments, we had to increase the iodine pressure and the laser power to obtain an absorption signal with sufficient intensity. This has introduced relatively large pressure and power broadening effects to the spectral linewidth. Furthermore, the linewidth of the diode laser ($2-4$ MHz) is much larger than that of the previous Nd:YAG laser (5 kHz). It is worth noting here that it is reasonable to increase the pressure and power broadening effects to the same level as the laser linewidth to realize the best figure of merit (linewidth $\times$ SN) for frequency stabilization using the observed spectrum. In terms of the signal, the signal intensity of the $R(36)32-0$ transition measured in the present work is expected to be similar to that of the previous $R(56)32-0$ transition \cite{Gerstenkorn1978}. However, a relatively large noise was observed on the baseline of the spectrum in our experiment. We found that this noise level was significantly reduced by tuning the laser frequency far from the iodine resonance. The large noise near resonance is considered to be because the large frequency noise (FM noise) of the diode laser is converted to amplitude noise (AM noise) by the absorptive response of the iodine. This FM-to-AM conversion process has been reported in previous studies using an atomic vapor (e.g., see Ref. \cite{Camparo2005}). It should be noted that this conversion effect is small for the narrow linewidth Nd:YAG laser. 

When the ground state of molecular iodine has an even (odd) rotational quantum number, the rovibrational energy level is split into 15 (21) sublevels, which results in 15 (21) hyperfine components. Frequency measurements of all the hyperfine components of rovibrational lines can contribute to a precise study of the hyperfine structures. A theoretical fit of the measured hyperfine splittings provides the hyperfine constants \cite{Borde1981}. The obtained hyperfine constants are important for improving or deriving formulas for the hyperfine interactions. For example, previous studies of the iodine hyperfine structures near 532 nm have revealed the rotation and vibration dependence of the hyperfine constants \cite{Hong2001JOSAB,Hong2002JOSAB}. Here we could in principle study the hyperfine structure of the observed $R(36)32-0$ transition. However, some of the hyperfine components near the center of the Doppler absorption were not observed. This is due to the strong linear absorption at the Doppler center, which degrades the signal level. In the near future, we plan to measure all the hyperfine components of the $R(36)32-0$ transition at lower iodine pressures using a much longer iodine cell. One alternative to the long iodine cell is an iodine-loaded hollow-core photonic crystal fiber \cite{Lurie2012}. The combination of the coin-sized laser module and the hollow-core photonic crystal fiber could be used to form an ultra compact iodine-stabilized laser. We note that with the new laser, wavelength conversion, and iodine devices, it is also attractive to investigate iodine transitions at other wavelengths \cite{Hong2009}.   

Our compact laser can be used for various applications including the interferometric measurement of gauge blocks \cite{Bitou2003}, the calibration of a wavelength meter, and as an absolute frequency marker for an astro-comb \cite{Wilken2012}. In the gauge block measurement, an uncertainty arises from the frequency modulation of a light source \cite{Bitou2003}. Therefore, we need to investigate the effect of the modulation frequency and width employed in the present work on the measurement precision of gauge blocks. Such experiments are now under way. For a wavelength meter, calibration is periodically needed, since its reading drifts due to variations in environmental temperature and pressure. For calibration purposes, a simple and low-cost laser with an accurate absolute frequency is suitable. An astro-comb has recently been demonstrated for the calibration of an astronomical telescope \cite{Wilken2012}. In this scheme, a frequency-stabilized laser is needed to identify the mode number of the comb. The laser must have a frequency accuracy of better than the repetition rate of the comb and be compact and robust enough for installation in an astronomical observatory.

\section*{Acknowledgments}
We are grateful to T. Suzuyama and M. Amemiya for maintaining UTC at NMIJ. We would like to thank Y. Bitou for discussions about the feasibility of the gauge block measurement.

\end{document}